\documentclass[5p]{elsarticle}

\usepackage[utf8]{inputenc}
\usepackage{float}
\usepackage{graphicx}
\usepackage{amssymb}
\usepackage{amsfonts}
\usepackage{amsmath}
\usepackage{booktabs}
\usepackage{tabu}
\usepackage{mathtools}
\usepackage{mathrsfs}
\usepackage{eurosym}
\usepackage{makecell}
\usepackage{tabularx}
\usepackage[bottom]{footmisc}
\usepackage{caption}
\usepackage{subcaption}
\usepackage{diagbox}
\usepackage{multirow}
\usepackage{color}
\usepackage[dvipsnames]{xcolor}
\usepackage{url}
\usepackage{sistyle}

\begin{document}
\sloppy

\title{Complementarity Assessment of South Greenland Katabatic Flows and West Europe Wind Regimes}

\author[ulgm]{David~Radu\corref{cor1}}
\ead{dcradu@uliege.be}
\author[ulgm]{Mathias~Berger}
\author[ulgm]{Raphaël~Fonteneau}
\author[ulgm]{Simon~Hardy}
\author[ulgg]{Xavier~Fettweis}
\author[rte]{Marc~Le~Du}
\author[rte]{Patrick~Panciatici}
\author[rte]{Lucian~Balea}
\author[ulgm]{Damien~Ernst}

\cortext[cor1]{Corresponding author.}
\address[ulgm]{Dept. of Electrical Engineering and Computer Science, University of Liège, Allée de la Découverte 10, 4000 Liège, Belgium}
\address[ulgg]{Laboratory of Climatology, Department of Geography, University of Liège, Belgium}
\address[rte]{R\&D Department, Réseau de Transport d'Electricité (RTE), France}

\begin{abstract}
Current global environmental challenges require vigorous and diverse actions in the energy sector. One solution that has recently attracted interest consists in harnessing high-quality variable renewable energy resources in remote locations, while using transmission links to transport the power to end users. In this context, a comparison of western European and Greenland wind regimes is proposed. By leveraging a regional atmospheric model specifically designed to accurately capture polar phenomena, local climatic features of southern Greenland are identified to be particularly conducive to extensive renewable electricity generation from wind. A methodology to assess how connecting remote locations to major demand centres would benefit the latter from a resource availability standpoint is introduced and applied to the aforementioned Europe-Greenland case study, showing superior and complementary wind generation potential in the considered region of Greenland with respect to selected European sites.

\end{abstract}


\maketitle


\section{Introduction}
\label{sec:intro}
A current trend in the power system community addresses renewable energy harvesting in remote, yet resource-rich locations and their subsequent integration via large-scale interconnections. In a future power system context defined by dominant variable renewable energy (VRE) generation and increased shares of electrical loads, linking separate power systems offers benefits on various operational levels. From a generation standpoint, one may see the potential of VRE harnessing in resourceful sites and subsequent delivery to major load centres via adequate transmission links. In addition, the negative impact that high VRE generation intermittency has on the operation of power systems could be reduced effectively through complementary production profiles originated from different resource patterns induced by time zone differences (on various latitudes), the timing of seasons (across longitudes) or distinct meteorological dynamics. From a load perspective, exploiting shifted consumption patterns between regions arising from the geographical positioning of the consumers at different longitudes and latitudes has the potential to level out aggregated load profiles. These would, in turn, lead towards a less challenging operation of power systems, together with a reduction in operational and planning costs.

Coupling distinct power systems from a country, to a regional and ultimately an intercontinental level would result in a globally interconnected electricity network, or a ``global grid''. The idea of a global grid was first proposed in \cite{gg1}, where the authors envision VRE technologies as crucial in meeting the ever-increasing electricity demand, with high-capacity interconnections being the backbone of the corresponding transmission infrastructure. The same article also describes various operational opportunities emerging from such a large-scale project and it highlights regulatory hurdles likely to arise in such a complex set-up. A more comprehensive, yet still conceptual view on the topic is provided in \cite{gg2}. The book provides a more detailed assessment of the motivation behind the development of a global grid before mapping specific regions for energy harvesting and routes for long-haul interconnections, and finally discussing the technical innovation required for the successful deployment of this project. Also, an economic dispatch model is the subject of ongoing work for a CIGRE Working Group \cite{gg3} that is investigating the technical feasibility of a global grid, as well as its economic competitiveness by assessing the trade-off between the cost of interconnectors and the benefits associated with remote VRE harvesting.

In the context of a global electricity interconnection, the scope of this paper is to assess the wind resource complementarity between two adjacent macro-regions as part of the broader global grid: Europe and Greenland. Wind availability in the former is sometimes an issue that leads to increased utilization of balancing units (usually fossil fuel-based generators) and storage capabilities. In this regard, seasonal patterns generally show inferior resource availability during summer time \cite{wins}, while winter conditions could also display wind resource scarcity coupled with usually low solar irradiance. A resource-rich and load-free region such as Greenland has the potential to provide wind energy to European users in times of local scarcity, while mitigating the balancing and storage requirements.

For the remainder of the paper, Section \ref{sec:worls} documents previous works related to remote VRE harvesting and resource potential assessment. Section \ref{sec:data} introduces the sources of wind data and briefly discusses local features of wind regimes in Greenland (i.e., katabatic winds) that are favourable for extensive VRE generation. Locations for wind power generation to be investigated are selected in Section \ref{sec:region_selection}. Section \ref{sec:methodology} details the methodology proposed to study the resource complementarity before results of the Europe-Greenland case study are presented in Section \ref{sec:results}. Finally, Section \ref{sec:conclusion} concludes the paper and proposes related future research directions.

\section{Related Works}
\label{sec:worls}

Harnessing renewable energy sources (RES) in remote locations to supply major load centres has long been seen as a way of achieving deep decarbonization goals for power systems located in areas scarce of renewable energy potential. One of the first projects of this kind revolves around the idea of supplying Europe with solar and wind energy originated from the VRE-abundant North African and Middle Eastern (MENA) territories \cite{erdle2010desertec, dii2012mena, dii2013mena, dii2014mena, platzer2016supergrid}. The first initiative in this direction was the DESERTEC project, which emphasized the vast potential for solar power generation in the MENA region that could account for 15\% of the electricity demand in Europe by 2050 \cite{erdle2010desertec}. A comprehensive three-part study published by a consortium of European and MENA (EUMENA) stakeholders further investigated this topic. Potential synergies arising from the integrated planning and operation of renewable-based EUMENA power systems are initially identified \cite{dii2012mena}, before possible economic and regulatory frameworks, as well as expected economic impacts are discussed \cite{dii2013mena}. Finally, a techno-economic study of potential routes for transmission corridors is proposed \cite{dii2014mena}. Furthermore, a geographic information system (GIS)-based analysis investigates the technical potential of wind, photovoltaic (PV) and concentrated solar power (CSP) in the Mediterranean region to cover the full electricity demand of Europe and North Africa by 2050 \cite{platzer2016supergrid}. Still in Europe, other initiatives have studied the prospect of harnessing renewable resources (e.g., wind, geothermal, hydro) in North-Western Europe for subsequent delivery to European demand centres \cite{3e2015northsea, skuli2014icelink}. Moreover, a recent study focused on estimating the potential of wind energy over open oceans concludes that, with adequate technological development, the North Atlantic area alone could power the entire world via wind converters \cite{possner2017oceans}. A similar approach to the one followed in the EUMENA studies is found in the Gobitec concept, which investigates the potential of harvesting VRE in the resource-rich Gobi Desert and delivering it to major load centres in East Asia. The report estimates the cumulated potential of the Gobi Desert in terms of wind and PV installed capacity at \SI{2600}{GW} \cite{fraunhofer2014gobitec}. The International Energy Agency (IEA) also documents the theoretical potential of solar power generation via large-scale PV installations in various regions known for their particularly high solar irradiation \cite{iea2015desert}. A more recent article investigates the potential for VRE harvesting in Australia and potential synergies between different energy vectors (e.g., electricity, gas, heat) to supply major demand centres in East and South-East Asia \cite{gulagi2017australia}. Another study, this time reporting on renewable energy resource distribution and quality in the North American region, reveals untapped geothermal, wind, hydro and tidal potential in Alaska and proposes several pathways to integrate it, including its transmission to load centres on the continent \cite{nrel2012alaska}.

More particularly, Greenland has also been the subject of VRE resource analysis for power generation. A first PhD thesis on this topic investigates the potential of wind power generation in Greenland by using a mesoscale atmospheric model to recreate local wind regimes \cite{dasoares2016uppsala}. Certain locations are selected for large-scale wind turbine (e.g., \SI{3}{MW} units) deployment and the study concludes that, even though the site selection process is highly complex, there is undisputed potential for wind power generation in Greenland. A second PhD thesis on the same subject combines micro- and mesoscale climate modelling for an accurate representation of local wind circulation \cite{wigr1}. The conclusion of the study supports the resource potential of Greenland for wind generation, with specific features of local wind regimes (e.g., semi-permanent occurrence of katabatic flows) found to facilitate increased levels of electricity generation. A work authored by a Nordic consortium also studies the potential of RES (e.g., hydro, wind, PV) in Greenland and different interconnection possibilities between the latter and northern Europe \cite{orku2016northatlantic}. In addition, the author of \cite{zhenya2016global} envisions Arctic regions (e.g., Greenland, Norwegian Sea, Barents Sea) as a next step of wind generation deployment in the North Sea, with a cumulated potential of electricity delivery to Europe and North America estimated at \SI{1800}{TWh} per year.

In addition to assessing regional VRE resources in terms of electricity generation potential, the current work follows the approach proposed in \cite{crit_time_wind}, where inter-regional VRE resource complementarity in both space and time is investigated by means of a parametrised family of scalar indicators. Moreover, the wind resource assessment in Greenland is conducted via a mesoscale climate model proven to accurately replicate wind circulation in polar regions \cite{mar}.

\section{Reanalysis Data and Katabatic Winds}
\label{sec:data}

The process of wind resource assessment, as proposed in this paper, starts with data acquisition. In this regard, collection of wind signals in Europe and Greenland at hourly resolution and covering ten years (i.e., 2008-2017) is achieved via two different sources. The first source, used for data collection in Europe, is the state-of-the-art ERA5 reanalysis \cite{era5} developed by the European Centre for Medium-Range Weather Forecasts (ECMWF) through the  Copernicus Climate Change Service (C3S). It is an atmospheric reanalysis model\footnote{Reanalysis is the process of using a data assimilation system (i.e., a sequential procedure in which model states are updated on-line while previous forecasts are continuously compared to available measurements) providing ``a consistent reprocessing of meteorological observations'' \cite{merra}.} that incorporates in situ and satellite observations at high temporal (i.e., down to hourly) and spatial (i.e., $0.28^{\circ}\times0.28^{\circ}$) resolution, at various pressure levels and currently covering the last 40 years (i.e., 1979 - present). Within the scope of the current paper, the ERA5 data used here is provided at a height of 100 meters above ground level and the hourly sampling rate chosen for wind potential assessment is achieved via linear interpolation from three-hourly output snapshots. The limitations of reanalysis models in estimating wind energy potential are reported in the particular case of another such model (i.e., MERRA2) used in the European context, with significant spatial bias being identified for specific sub-regions \cite{iain}, partly resulting from the coarse spatial resolution used to model the local or topography-induced winds. A comparison between the two reanalysis models (i.e., ERA5 and MERRA2) \cite{re_comp} concludes that such tools are not recommended for estimating mean wind speeds for given locations due to their limitations in solving ``local variations, especially in more complex terrain''. 

\begin{figure}[b]
	\centering
	\includegraphics[width=0.49\textwidth]{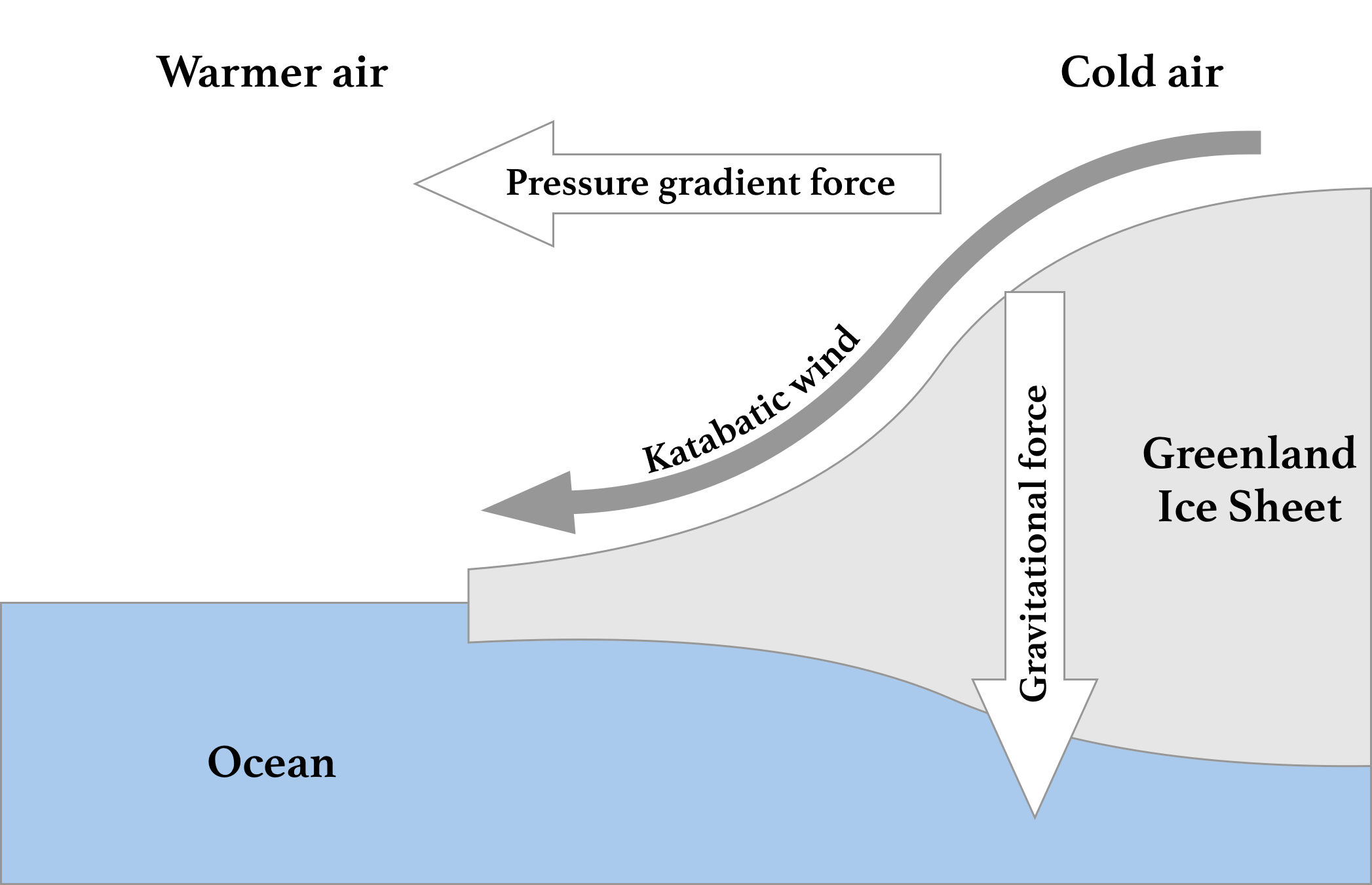}
	\caption{An illustration of katabatic winds in Greenland, carrying high-density air from a higher elevation down a slope under the force of gravity. Adapted from \cite{hgrobe}.}
	\label{fig:katabatic_wind}
\end{figure}

In order to overcome the limitations of the aforementioned tools when investigating the wind generation potential of Greenland, wind signals in this particular region are retrieved from a second source, i.e., the regional MAR (Modèle Atmosphérique Régional) model. MAR is a climate model developed specifically for simulating climatic conditions of polar regions and has been repeatedly validated over Greenland \cite{mar}. MAR, as an atmospheric model\footnote{An atmospheric model is a mathematical model based on a set of dynamical equations governing atmospheric motions and using numerical methods to obtain approximate solutions of the studied system of coupled equations.}, solves a set of dynamical equations over a limited integration domain by using reanalysis-based fields (here coming from ERA5) as lateral boundary conditions (e.g., temperature, wind, humidity, pressure at each vertical level of the MAR model). The choice of MAR for estimating Greenland's wind potential is based on its specific ability to accurately represent, at higher resolution (down to \SI{5}{km}$\times$\SI{5}{km}), physical processes in Greenlandic regions, including the local, gravity-driven katabatic winds. In this work, hourly values of wind speed at 100 meters above ground level are generated using MAR.

The most promising, yet underestimated source of wind generation potential in Greenland stems from the existence of katabatic flows. These local atmospheric movements are the result of heat transfer processes between the cold ice cap and the warmer air mass above it. In brief, when the air mass temperature is higher than that of the ice sheet, the former is cooled down by radiation, thus the air density increases forcing it down the sloping terrain, as depicted in Figure \ref{fig:katabatic_wind}. The flow of katabatic winds is driven by gravity, temperature gradient and inclination of the slope of the ice sheet \cite{dasoares2016uppsala}. This wind develops in the first tens of meters above surface (in the boundary layer) with a relatively constant direction down the slope of the terrain, is quasi-constant, but is strengthened when an atmospheric low-pressure area approaches the coast. Katabatic winds develop on a daily basis, regardless of the season, with a slight diurnal shift in their occurrence according to the season (i.e., arrival at the edge of the ice cap during early mornings throughout the winter, around noon during the summer). In addition, the highest intensity of katabatic winds is reported to occur on the south-eastern coast of Greenland, mainly due to characteristic steep slopes and flow-channelling conditions \cite{wigr1}.

\section{Region Selection}
\label{sec:region_selection}

\begin{figure*}
	\begin{subfigure}[b]{0.32\textwidth}
		\includegraphics[width=\textwidth]{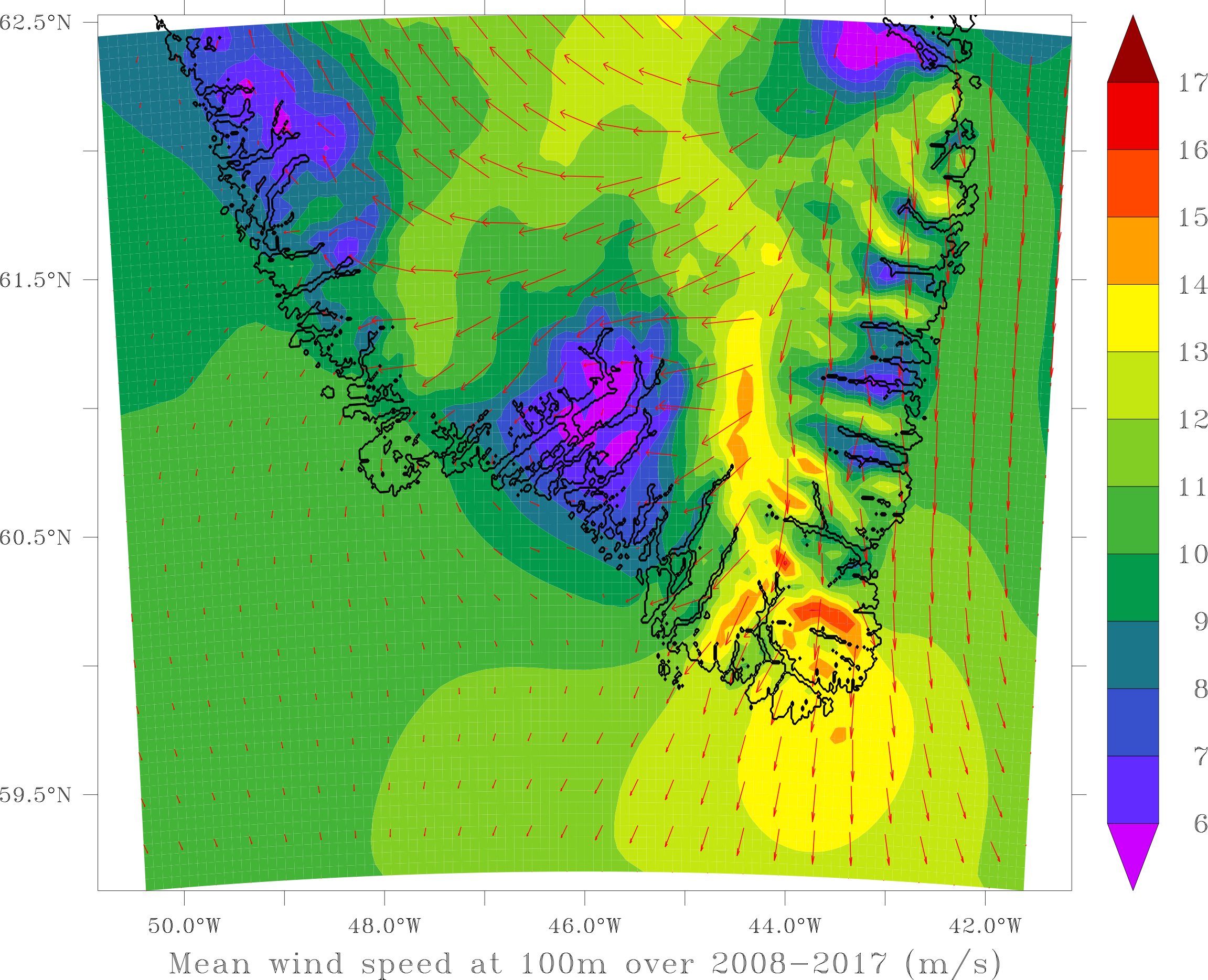}
		\caption{}
		\label{fig:greenland_s_full}
	\end{subfigure}
	\begin{subfigure}[b]{0.32\textwidth}
		\includegraphics[width=\textwidth]{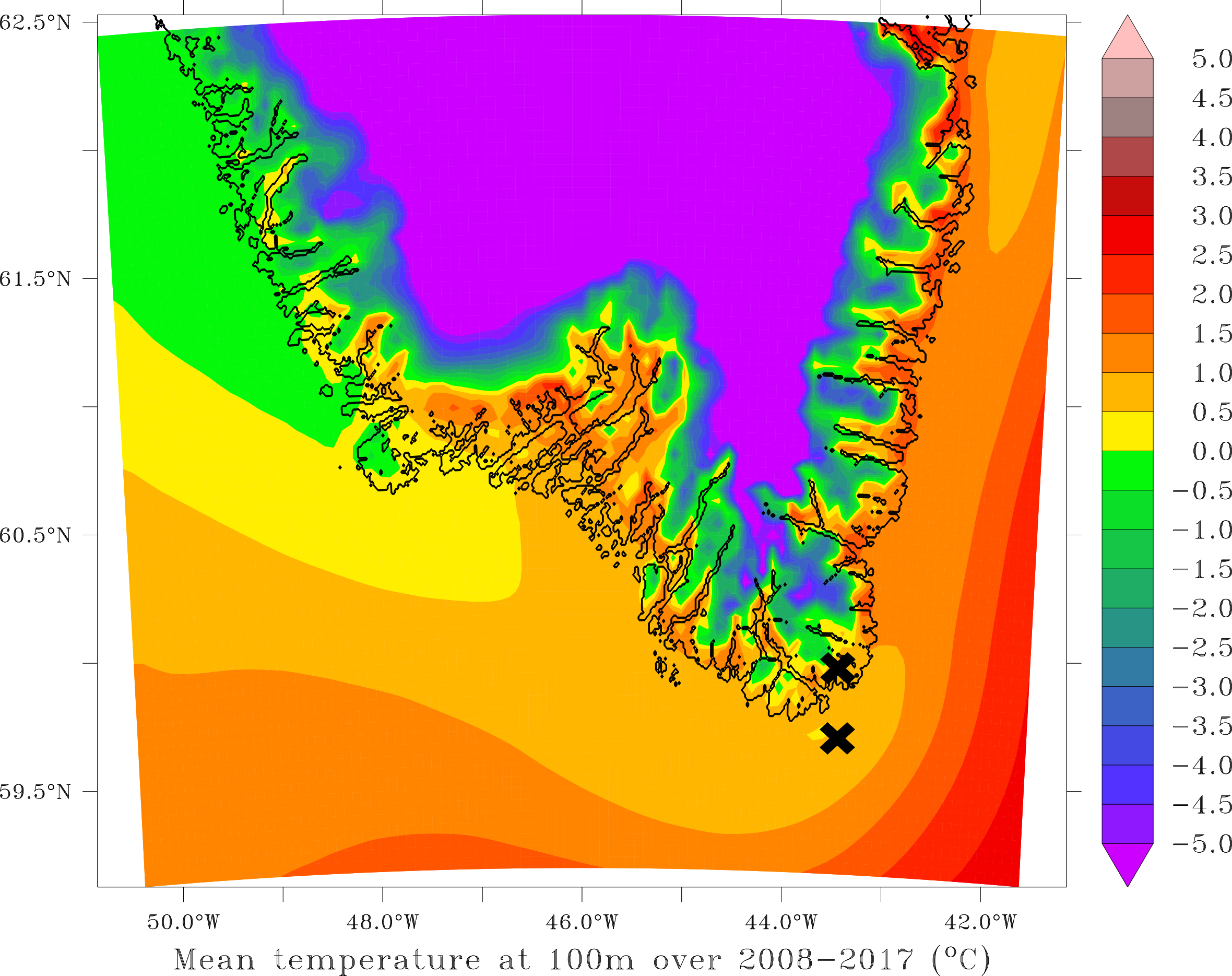}
		\caption{}
		\label{fig:greenland_s_temp}
	\end{subfigure}
	\begin{subfigure}[b]{0.32\textwidth}
		\includegraphics[width=\textwidth]{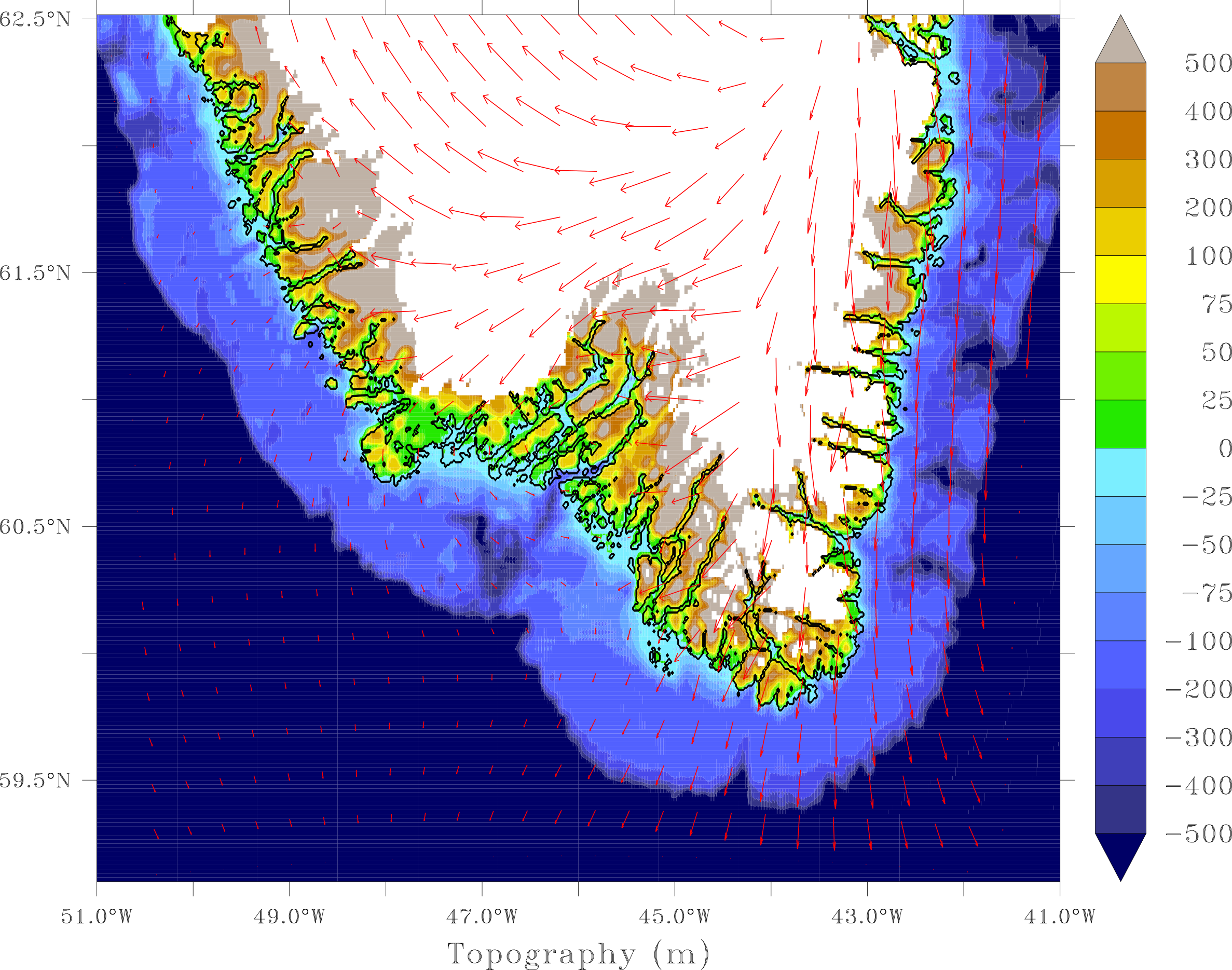}
		\caption{}
		\label{fig:greenland_s_ice}
	\end{subfigure}
	\caption{(a) Greenland average wind speed magnitudes (m/s) as provided by MAR for 2008-2017. The contour lines are built from average wind speeds at a height of \SI{50}{m} above ground level, with a spatial resolution of \SI{5}{km}$\times$\SI{5}{km}. (b) South Greenland average temperature profiles as computed with MAR for 2008-2017. The contours are constructed from annual mean temperature in $^\circ$C at \SI{100}{m} above ground level, with a spatial resolution of \SI{5}{km}$\times$\SI{5}{km}. (c) South Greenland topography superimposed over the land area not covered by permanent ice. The data is expressed in metres and has a spatial resolution of \SI{1}{km}$\times$\SI{1}{km}.}
\end{figure*}

Site selection in Greenland relies on a visual inspection process of the local wind regimes. As seen in Figure \ref{fig:greenland_s_full}, Greenland's southernmost region is the most promising from a wind resource perspective, therefore selection of the assessment point is constrained within the yellow and red-coloured areas plotted on the chart, ones with modelled average wind speeds above 13 m/s. In fact, availability of such high average wind speeds is the consequence of the common direction of the general circulation driven winds (as shown on the same chart) and the local katabatic winds that prevents the two atmospheric motions from cancelling each other out. Selection of an onshore point (i.e., GR\textsubscript{on}) in this area of interest is further supported by year-long high temperatures (associated with low icing risks wind turbine components - Figure \ref{fig:greenland_s_temp}) and the absence of a permanent ice sheet, as well as by the characteristic low elevation (Figure \ref{fig:greenland_s_ice}). In addition, an offshore location (i.e., GR\textsubscript{off}), just south from the onshore one, will be assessed. The choice of the latter location is also supported by the bathymetry of Greenland's territorial waters, with depths below 100 metres. The two sites are marked with a black marker in Figure \ref{fig:greenland_s_temp}.

\begin{figure}
	\centering
	\includegraphics[width=0.49\textwidth]{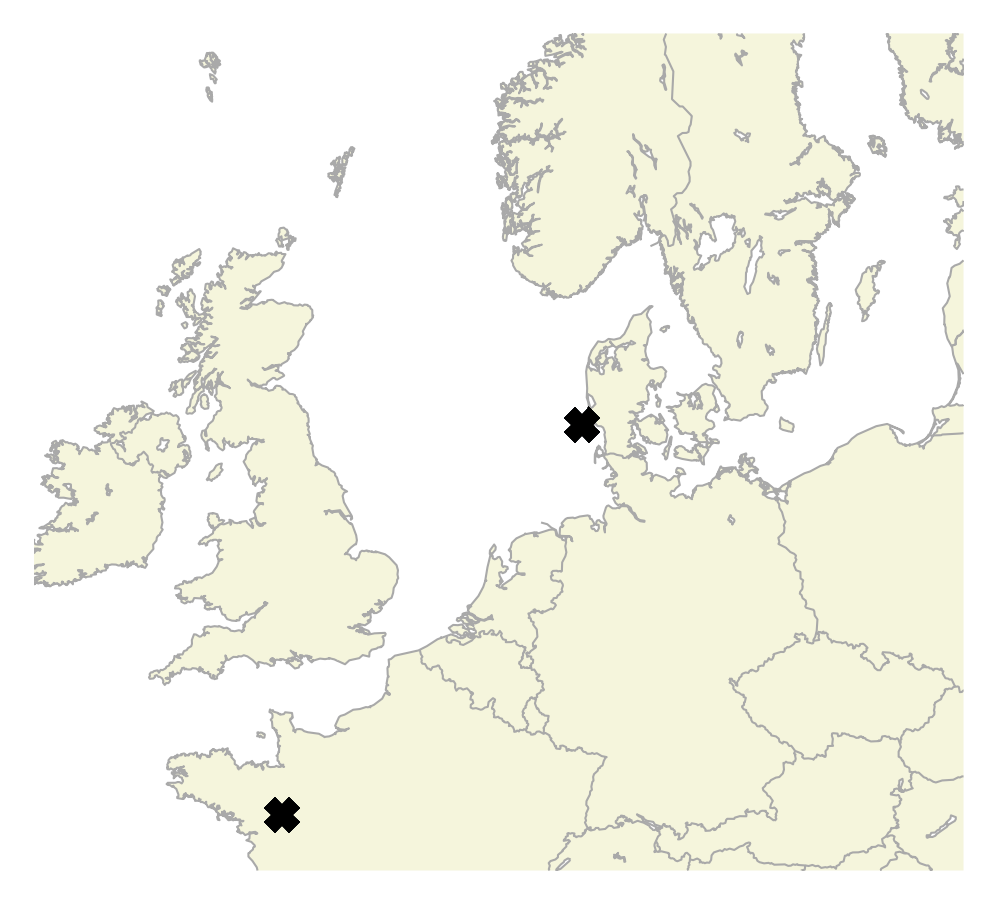}
	\caption{Location of the two European wind farms investigated.}
	\label{fig:wind_europe}
\end{figure}

Selection of the European generation sites to be compared with the locations in Greenland is initially bound to the region adjacent to or within the North Sea basin, one of the most productive areas on the continent \cite{wind_ewea}. Within these boundaries, two locations are selected based on the existence of operational wind farms. More specifically, the selected points coincide with the geographical coordinates of the Horns Rev (Danish offshore) - DK - and Portes de Bretagne (French onshore) - FR - wind farms. The location of the two wind farms is depicted with a black marker in Figure \ref{fig:wind_europe}.

\section{Methodology}
\label{sec:methodology}

In this section, starting from a basic power generation metric (i.e., the hourly average capacity factor), we carry on with the assessment of wind resource by exploiting the notions of complementarity and critical time windows \cite{crit_time_wind} for assessing to which extent aggregating geographical locations may decrease the occurrence of low-generation time periods on a system-wide basis.

Let $l = (\lambda^{lon}, \lambda^{lat}) \in \mathbb{R}^2$ denote an arbitrary geographical location given by its longitude and latitude, and $\mathcal{L} \subset \mathbb{R}^2$ be the set of all locations of interest. Then, let $\tau \subset \mathbb{R}$ be a continuous time horizon over which signals are recorded. Additionally, let $T \in \mathbb{N}$ denote a number of time steps. The continuous time horizon $\tau$ is discretised into a set of $T$ discrete time instants $\mathcal{T} =\{t_0, \ldots, t_{T-1}\}$, with $t_{k+1} = t_{k} + \Delta, \mbox{ } \forall k \in [0, T-1] \cap \mathbb{N}$, and $\Delta = (\mbox{max}\{\tau\} - \mbox{min}\{\tau\})/(T-1)$. Moreover, let $\mathbf{x}^l \in \mathcal{R}^T$ be a vector representing a signal which takes values in a set $\mathcal{R}$, has been recorded at some location $l \in \mathcal{L}$ over some time duration $\tau$ and (uniformly) sampled with period $\Delta$. Each sample of this time series will be denoted as $x_t^l \in \mathcal{R}, \mbox{ } \forall t \in \mathcal{T}$. For instance, it may be a time series of hourly wind speeds or capacity factors, in which case $\mathcal{R} = \mathbb{R}_{\ge 0}$ and $\mathcal{R} = [0, 1]$, respectively.

\vspace{0.25cm}
\subsection{Average Capacity Factors}
\label{subsub:average_load_factor}

Let $F : \mathbb{R} \mapsto [0,1]$ denote a transfer function associated with a given wind turbine technology. This function maps an hourly average capacity factor for any wind input signal $s_t^l \in \mathbb{R}_{\ge 0}$. We also define by $\mathbf {u}^{l} \in [0,1]^{T}$ the capacity factor time series, such that the following component-wise relationship holds
\begin{equation}
u^{l}_t = F \left( s^{l}_t \right) \hspace{1mm}, \forall \hspace{1mm} l \in \mathcal L, \forall \hspace{1mm} t \in \mathcal T.
\end{equation}
Finally, the average capacity factor for a given location can be expressed as
\begin{equation}
e^{l} = \frac{1}{T}\sum_{t=0}^{T-1} u^{l}_t, \quad \forall \hspace{1mm} l \in \mathcal L \ .
\end{equation}

\begin{figure}[!b]
	\centering
	\includegraphics[width=0.49\textwidth]{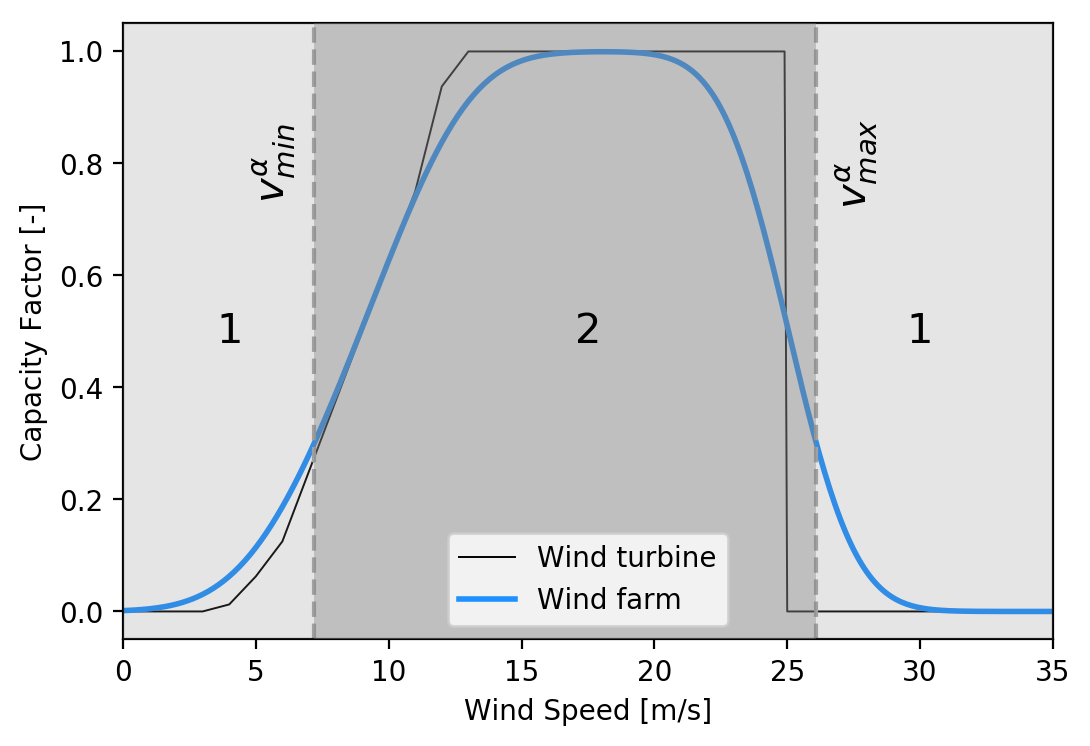}
	\caption{Single turbine and wind farm transfer functions. Example of wind farm power curve aggregation based on multiple \textit{aerodyn SCD 8.0/168} units. Power output regimes $(1, 2)$ with the two wind speed thresholds - $v_{min}^{\alpha}, v_{max}^{\alpha}$ - associated with a capacity factor threshold ($\alpha$) of 30\% also displayed.}
	\label{fig:pcurve}
\end{figure}

The selection of an appropriate transfer function $F$ is based on a multi-turbine power curve approach proposed in \cite{iain} and also used in \cite{crit_time_wind}. In this regard, we make use of an aggregated transfer function modelled via a Gaussian filter (depicted in Figure \ref{fig:pcurve}) that emulates the dynamics of a wind farm comprised of identical individual units, while taking as input the wind signal of one single point within this farm.

\vspace{0.25cm}
\subsection{Resource Complementarity}
\label{subsub:complementarity_cf_basis}

For any location $l \in \mathcal{L}$, let $\mathcal{B}^l = \{b_0^l, \ldots, b_N^l\} \subset \mathbb{R}^{N+1} \cup \{-\infty, \infty\}$ be a finite set whose elements are such that $b_{n-1}^l < b_n^l, \mbox{ } \forall n \in [1, N] \cap \mathbb{N}$, and from which a set of intervals $\mathcal{I}^l = \{I_1^l, \ldots, I_{N}^l\}$ can be constructed, such that $I_n^l = [b_{n-1}^l, b_{n}^l] \subset \mathbb{R}, \mbox{ } \forall n \in [1, N] \cap \mathbb{N}$ and $\cup_n I_n = \mathcal{R}$. Furthermore, a set of labels $\mathcal{C}^l = \{c_1^l, \ldots, c_N^l\} \subset \mathbb{N}$ is also assigned to a each location $l \in \mathcal{L}$. The elements and cardinality of $\mathcal{B}^l$ and $\mathcal{C}^l$ may change from one location to the next, but for any location $l \in \mathcal{L}$, one always has $|\mathcal{C}^l| = |\mathcal{I}^l| < \infty$. Then, let $h^l: \mathcal{R} \mapsto \mathcal{C}^l$ be a family of mappings such that 
\begin{equation}
h^l(x) = \left \{ \begin{array}{cc} c_1^l &\mbox{ if } x \in I_1^l\\
 & \vdots \\
c_N^l &\mbox{ if } x \in I_N^l \end{array} \right . .
\end{equation}
One such mapping is defined for each location $l \in \mathcal{L}$ and can be used to cluster any input time series taking values in $\mathcal{R}$ into discrete classes. Now, let $g_{nm}:\mathbb{N} \times \mathbb{N} \mapsto \{0,1\}$ be a family of mappings associating a binary digit value to a pair of labels, such that
\begin{equation}
g_{nm}(c_n,c_m) =
\begin{cases} 
1, & \mbox{ if } \hspace{5pt} (c_n,c_m)=(n,m) \\
0, & \mbox{ otherwise} \\
\end{cases}.
\end{equation}
Then, for any two locations $(l_1,l_2) \in \mathcal L \times \mathcal L$, associated label sets $\mathcal{C}^{l_1}$, $\mathcal{C}^{l_2}$ and signals $\mathbf{x}^{l_1}, \mbox{ } \mathbf{x}^{l_2} \in \mathcal{R}^T$, one can construct a matrix $C^{(l_1,l_2)} \in [0, 1]^{|\mathcal{C}^{l_1}| \times |\mathcal{C}^{l_2}|}$ with entries:

\begin{equation}\label{eq:cfactors}
C_{nm}^{(l_1,l_2)} = \frac{1}{T} \sum_{t=0}^{T-1} g_{nm} \bigg({h^{l_1}(x^{l_1}_t), h^{l_2}(x^{l_2}_t)\bigg)},
\end{equation}
where it is understood that $x_t^{l_1}$ and $x_t^{l_2}$ represent the entries of $\mathbf{x}^{l_1}$ and $\mathbf{x}^{l_2}$, respectively. In the sequel, for the sake of clarity, the superscripts $(l_1, l_2)$ will be dropped when writing the entries of $C^{(l_1,l_2)}$. Moreover, the coefficients $C_{nm}$ that may be computed as given in Equation \ref{eq:cfactors} will be referred to as complementarity factors in what follows. Put simply, the complementarity factor $C_{nm}$ quantifies how often the signal observed at location $l_1$ takes values corresponding to class $n$, whilst the signal recorded at location $l_2$ takes values associated with class $m$. In general, one therefore has $C_{nm} \ne C_{mn}$.

In the complementarity analysis proposed in this work, the underlying signal represents the hourly average capacity factors, while the associated classes correspond to low and high power generation regimes, respectively. It should be mentioned at this point that complementarity should not be understood in the usual sense of correlation (as computed on detrended signals via standard measures, such as Pearson, Spearman or Kendall correlation coefficients), but rather as the assessment of situations in which system-side, low-generation events occur, a detrimental feature of power systems characterized by high shares of VRE generation. Roughly speaking, in terms of complementarity factors, such behaviour would translate into high $C_{nm}$ values for classes $n$ and $m$ associated with low power generation regimes. Thus, in later developments, signals will be considered complementary if $C_{nm}$ values associated with low production regimes are small.

\vspace{0.25cm}
\subsection{Critical Time Windows}
\label{subsub:critical_time_window_occurence}

Given a time duration $\delta \in \{1, \ldots, T\}$, we define a  time window $w^{\delta}_t$ as being a set of $\delta$ integers starting at time $t$
\begin{equation}
w^{\delta}_t = [t, t+\delta - 1] \cap \mathbb  N \ .
\end{equation}
In addition, the set of all $\delta$-time windows within a time domain that starts at $T_s$ and ends at $T_f > T_s$, such that $(T_f-T_s \ge \delta)$ is denoted as $\mathcal W^{\delta}$ and can be defined as
\begin{eqnarray}
\mathcal W^{\delta} = \left\{ w^{\delta}_t | t \in \{T_s, \ldots, T_f-\delta \} \right\} \ .
\end{eqnarray}
Note that $\mathcal W^{\delta}$ is a set of sets of integers. Also, we introduce a mapping ${\hat {U}_\delta} : \mathcal W^\delta \times [0,1]^T \mapsto [0,1]^{\delta}$ dedicated to extracting a $\delta$-length truncation of a capacity factor time series $\mathbf u^l$ over a time window $w_t^\delta$
\begin{eqnarray}
\hat{U}_{\delta} \Big({w^{\delta}_t}, \mathbf{u}^{l} \Big) = \left[ u^{l}_t , u^{l}_{t+1}, \ldots, u^{l}_{t+\delta-1} \right].
\end{eqnarray}
Then, we define a mapping $N_\delta:[0,1]^\delta \mapsto [0,1]$ that returns a scalar between zero and one from an input vector of appropriate dimensions, as follows
\begin{eqnarray}
N_{\delta}({\bold v}) = \frac{1}{\delta} \sum_{i=1}^{\delta} v_i \ .
\end{eqnarray}
Standard statistical indicators (e.g., a given quantile) can be straightforwardly integrated in this mapping. In the case at hand, $N_{\delta}({\bold v})$ represents the average value of the vector ${\bold v}(i)$ over its time domain.
Let $\alpha \in [0,1]$ be a capacity factor threshold. For a given location $l \in \mathcal L$ , we denote by $\Omega^{l}_{\delta \alpha}$ the set of $(\delta, \alpha)$-critical time windows that gathers all $\delta$-time windows during which the $N_\delta$ measure is smaller than $\alpha$
\begin{eqnarray}
\Omega^{l}_{\delta  \alpha} = \left\{  w_t^\delta \bigg|  w_t^\delta \in \mathcal W^{\delta},   N_{\delta}\Big(\hat{U}_{\delta} \Big({w^{\delta}_t}, \mathbf{u}^{l} \Big)\Big) \leq \alpha \right\} \ .
\end{eqnarray}
Let $ L \in \mathcal P(\mathcal L)$ be a non-empty subset of locations. We introduce the set $\xi_{\delta \alpha}(L)$ as being the intersection of the sets of $(\delta,  \alpha)$-critical time windows over the subset of locations $L$ 
\begin{eqnarray}
\forall L \in  \mathcal P(\mathcal L), |L| \ne 0, \hspace{1mm}
\xi_{\delta \alpha}(L) = \bigcap_{l \in L} \Omega^{l}_{\delta \alpha} \ .
\end{eqnarray}
Intuitively, such a set contains the time windows simultaneously critical across all locations in $L$, that is, the time windows over which the average output power is smaller than or equal to $\alpha$. Finally, we define the last metric for wind resource assessment as a mapping $\Gamma_{\delta  \alpha} : \mathcal P(\mathcal L) \mapsto [0,1]$
\begin{eqnarray}
\forall \hspace{1mm} L \in  \mathcal P(\mathcal L), |L| \ne 0, \hspace{1mm} \Gamma_{\delta \alpha}(L) =  \frac{| \xi_{\delta \alpha}(L)|}{|\mathcal W^{\delta}|} \ .
\end{eqnarray}
Concretely, $\Gamma_{\delta \alpha}(L)$ is the proportion of $\delta$-time windows found to be critical at every location in $L$, according to the resource quality assessment criterion given by the measure mapping $N_\delta$.

\section{Results}
\label{sec:results}
\subsection{Wind Resource Assessment}
\label{sub:res_magnitude}

\begin{figure}
	\centering
	\includegraphics[width=0.49\textwidth]{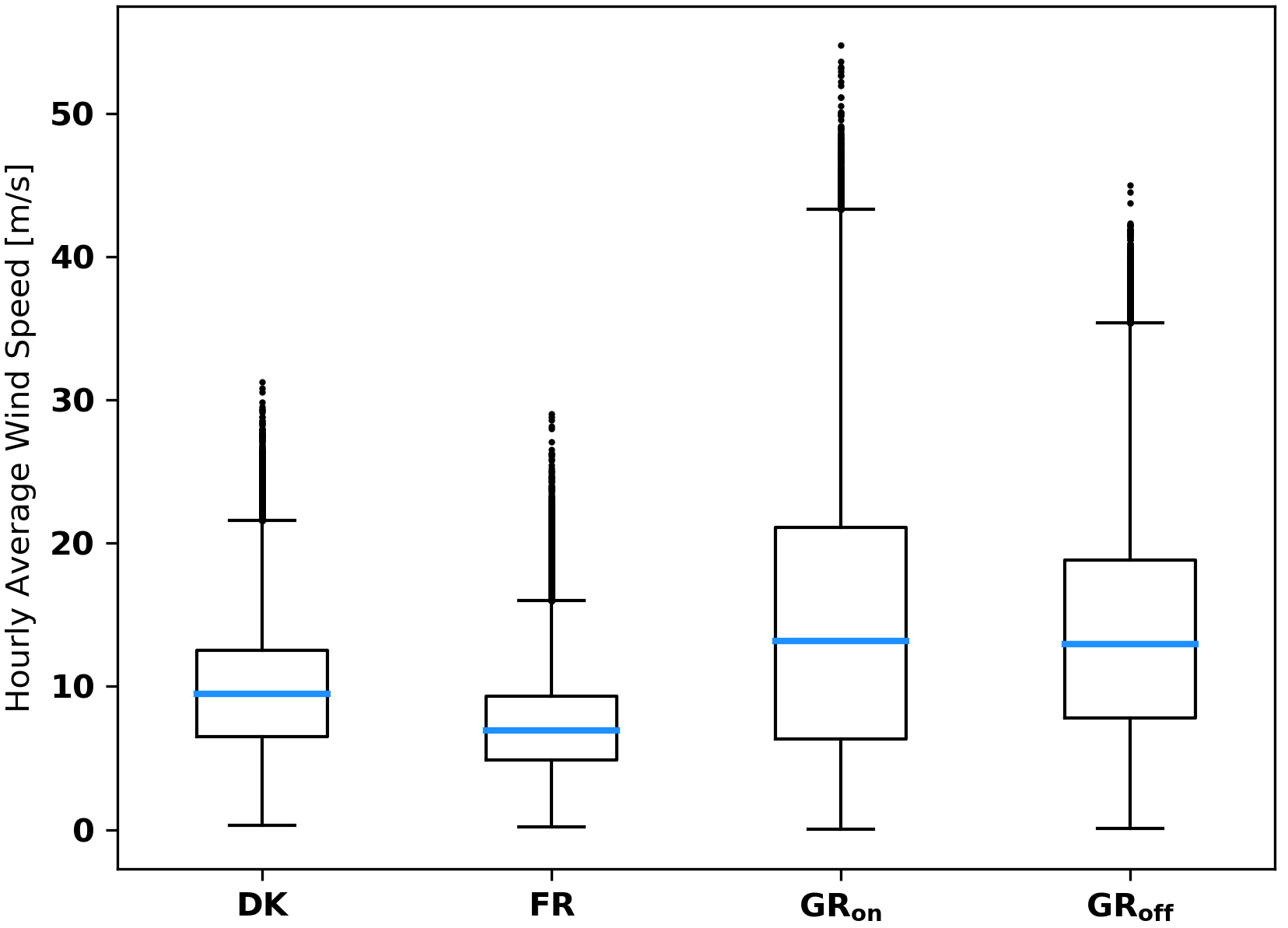}
	\caption{Boxplots providing descriptive statistics of wind signals for the four locations under consideration.}
	\label{fig:boxplot}
\end{figure}

The descriptive statistics of the wind time series associated with the studied locations are provided in Figure \ref{fig:boxplot}. The ten-year mean wind speed in both Greenlandic locations (i.e., around 14 m/s) is significantly higher than in both European sites (topped by the Danish offshore site, with an average wind speed of close to 10 m/s). In addition, a larger spread of modelled wind speeds in the Greenlandic regions can be observed. We note that, as reported in \cite{turb_wind}, a high standard deviation of the wind signals usually corresponds to increased turbulence intensity (i.e., short-term wind magnitude fluctuations relative to the mean speed) that may, in turn, negatively affect the performance of the wind farm. Nonetheless, it has been observed that larger standard deviation values corresponding to the sites in Greenland are not the result of short-term variations of the underlying wind signal, but are rather due to the strong influence of seasonality of the local natural resource, and may therefore not be associated with high turbulence intensities.

\begin{figure*}
	\centering
	\includegraphics[width=\textwidth]{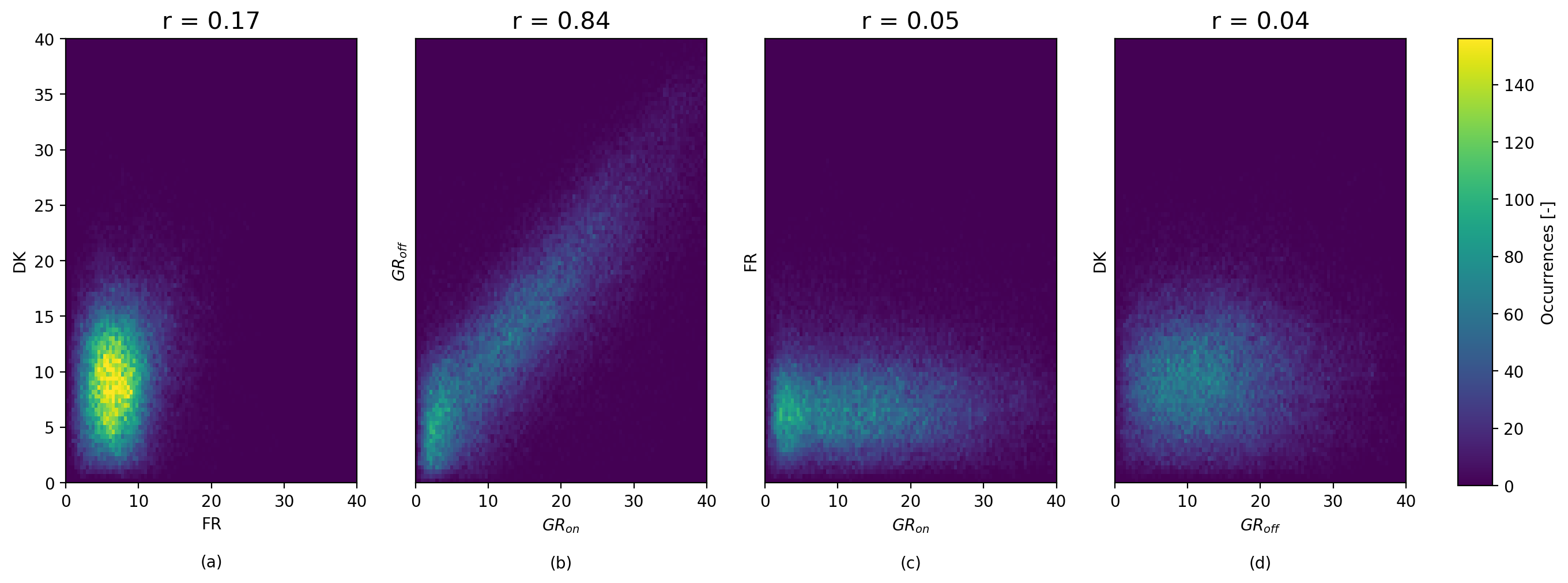}
	\caption{Bivariate histograms of wind signals for (a) the European, (b) Greenlandic, (c) the two onshore and (d) the two offshore locations. Each histogram bin corresponds to a \SI{0.5}{}$\times$\SI{0.5}{m/s} square.}
	\label{fig:histogram}
\end{figure*}

Bivariate histograms of wind speed time series are displayed in Figure \ref{fig:histogram} as a first indicator of resource complementarity associated with selected pairs of locations. The first plot (Figure \ref{fig:histogram}a) shows the approximate joint distribution of wind speeds in DK and FR (i.e., the European locations). Firstly, better wind resource at the former site is evident from the histogram, but high wind speeds (above 20 m/s) seldom occur in any of the two European locations. Then, a structured pattern featuring a very high concentration of data points between 5 and 10 m/s reveal a non-negligible degree of correlation between these sites. This analysis is further supported by a Pearson correlation coefficient (or $r$ index) value of 0.17, which, although modest, is much higher than that computed for pairs of remotely located sites, as discussed later. In Figure \ref{fig:histogram}b, a clear linear pattern is observed in the histogram, suggesting a large degree of correlation between wind regimes at the two Greenlandic locations (an expected outcome considering the close geographical proximity of the two locations). This claim is further backed by a Pearson correlation coefficient score of 0.84, by far the largest among all considered cases. The same analysis for pairs of onshore (FR-GR\textsubscript{on}) and offshore (DK-GR\textsubscript{off}) locations, respectively, is depicted in the last two subplots. The shape of the distribution in Figure \ref{fig:histogram}c reveals significantly superior resource at the Greenlandic onshore location compared to the European one, as well as very little correlation between wind signals (with an $r$ score of $0.05$). Regarding the offshore sites (Figure \ref{fig:histogram}d), slightly superior resource is observed in Greenland compared to the European offshore location. Moreover, a relatively wide-spread and even distribution of data points, especially for wind speeds between 5 and 20 m/s, suggests lack of correlation between signals, a feature supported by the associated $r$ index of $0.04$.

\subsection{Wind Farm Capacity Factor Comparison}
\label{sub:res_cfs}

\begin{table}
	\centering
	\renewcommand{\arraystretch}{1.2}
	\caption{Average capacity factors for the studied wind generation sites considering a transfer function associated with an aggregated wind farm for (i) a cut-out wind speed of \SI{25}{m/s} and (ii) an ideal cut-out wind speed superior to the maximum wind speed observed at different locations ($\underset{l,t}{max} \, s_{t}^{l}$).}
	\label{tab:cf}
	\begin{tabular}{c|cccc}
		$v_{cut}^{out}$ (m/s) & \textbf{DK} & \textbf{FR} & \textbf{GR\textsubscript{on}} & \textbf{GR\textsubscript{off}}  \\
		25 & 0.55 & 0.32 & 0.50 & 0.59 \\
		$\underset{l,t}{max} \, s_{t}^{l}$ & 0.56 & 0.33 & 0.66 & 0.69 \\
	\end{tabular}
\end{table}

Table \ref{tab:cf} shows estimated values for average capacity factors computed as proposed in Section \ref{subsub:average_load_factor}, assuming wind farm availability of 100\% (no losses due to icing, down times, etc.). Compared to available operational data, the average capacity factor of the European sites is inflated by approximately 10\%, assuming the currently in-use cut-out speed value of \SI{25}{m/s} \cite{dkwind}, \cite{frwind}. These overestimates were expected considering the 100\% availability assumption and the overestimation in reanalysis models of wind resource potential in northern and western Europe, as reported in \cite{iain}. Therefore, leveraging the recurrent validation of MAR in accurately replicating wind conditions in polar regions \cite{mar}, the differences between the capacity factors in the two Greenlandic locations and the ones associated with the European sites are even greater than those which can be inferred from Table \ref{tab:cf}. The second row of the same table shows the maximum theoretical capacity factor under the assumption that the individual units comprising a wind farm have a cut-out speed superior to any local wind speed to which they are exposed. Besides this assumption of an ``infinite'' cut-out speed, the transfer function of the wind farm remains otherwise the same. In other words, (i) capacity factor values are set to 1 for all wind speeds higher than that at which a capacity factor of 1 is first achieved and (ii) the low-wind, ramp-up regime of the wind farm is unaltered. In this case, while the average capacity factors of the European sites are barely affected (indicating very few occurrences of wind velocities above the current cut-out speeds), the same thing cannot be said about the locations in Greenland. There, under increased cut-out speed conditions, the onshore site would have the highest capacity factor gain (i.e., 16\%), while an offshore wind farm could reach capacity factors of almost 70\%.

\begin{figure}
	\centering
	\includegraphics[width=0.49\textwidth]{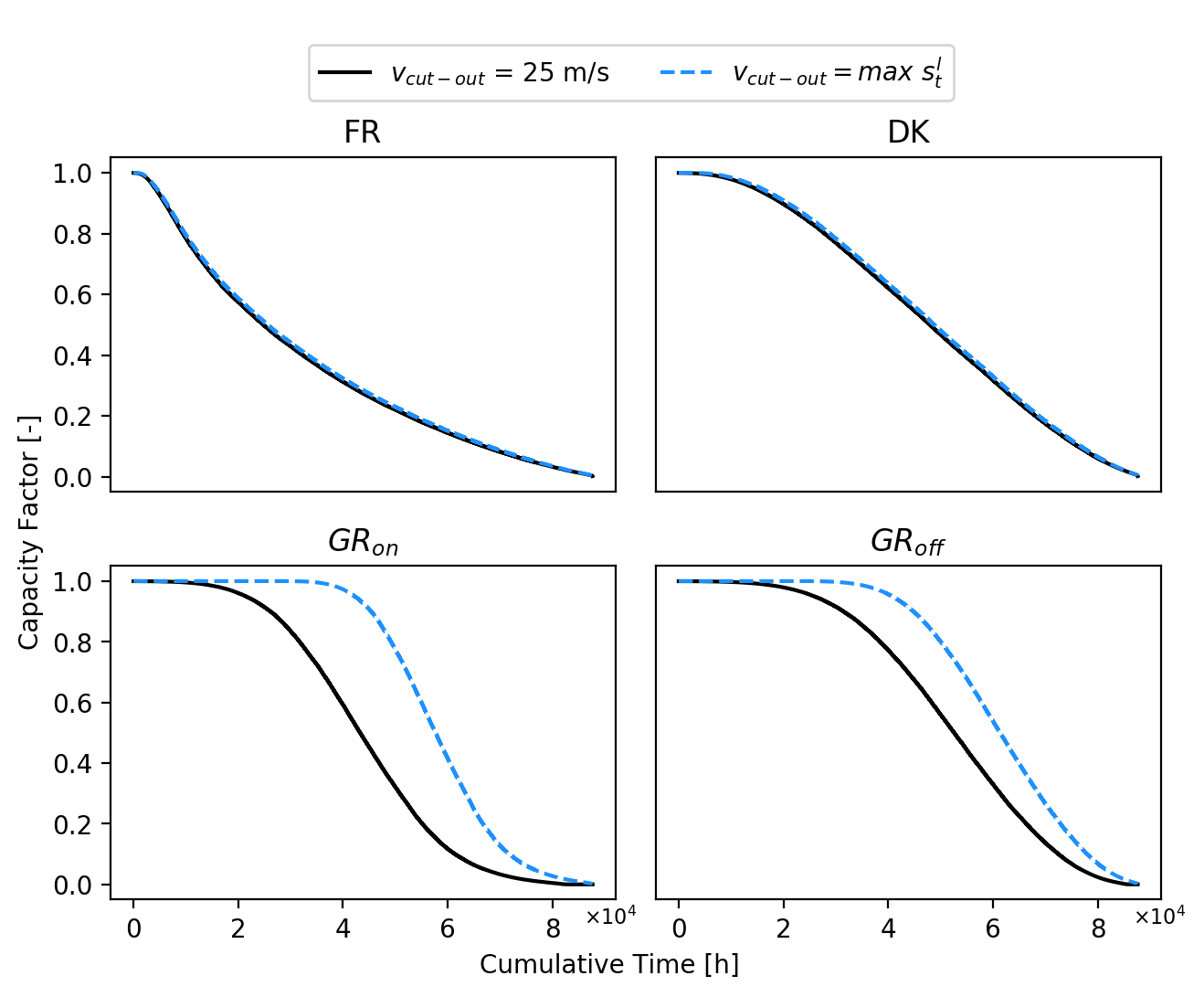}
	\caption{Duration curves of the four considered locations over the entire time horizon (2008-2017) assuming (i) a cut-out speed of individual wind converters of 25 m/s and (ii) an ideal cut-out speed superior to the maximum wind speed observed at each location.}
	\label{fig:duration}
\end{figure}

These findings are supported by the duration curves depicted in Figure \ref{fig:duration}. On the one hand, overlapping curves in the two subplots at the top reflect marginal gains in terms of wind farm output for the European locations, when technological development (i.e., increased cut-out speeds of wind converters) is assumed. On the other hand, assuming availability of wind converters with cut-out speeds above the maximum wind speeds of each location results in massive output improvements in Greenland. In fact, for both locations, capacity factors of 90\% or higher occur during more than half of the time. In this context, Figure \ref{fig:duration} clearly shows the lost potential of wind-based electricity generation in Greenland due to current technological limitations and indicates that novel wind turbine designs are required to fully harness the superior wind resource available in such regions.

\subsection{Potential of Wind Generation Complementarity}
\label{sub:wind_compl}

In line with the previously detailed methodology, power output complementarity factors for selected pairs of locations will be evaluated in the upcoming section. Two classes representing a low- and a high-generation regime, respectively, are defined by wind speed values associated with a given capacity factor threshold, for a particular conversion technology. Figure \ref{fig:pcurve} shows the separation of these classes via two wind speeds ($v_{min}^{\alpha}$ and $v_{max}^{\alpha}$) for a capacity factor threshold of 30\% and assuming the conversion technology introduced in Section \ref{subsub:average_load_factor}.

Power output complementarity factors for selected pairs of Greenlandic and European sites and for a capacity factor threshold of 30\% are displayed in Table \ref{tab:corr}. Each cell in these tables corresponds to a pair of capacity factor classes (as depicted in Figure \ref{fig:pcurve}) and a pair of locations, and contains information simultaneously recorded at each location and belonging to each corresponding class. Firstly, the aggregation of the two European locations (DK-FR) reveals a 17\% (resp. 34\%) probability of both sites yielding low (resp. high) output, while the two locations complement each other for 49\% of the time. Moreover, superior resource at DK is observed in the complementarity factors associated with different output regimes (the probability of high output at DK occurring simultaneously with low output at FR is 36\%, while the opposite situation happens only 13\% of the time). Secondly, considering the aggregation of both locations in Greenland (GR$\textsubscript{on}$-GR$\textsubscript{off}$), we observe a fairly high probability (74\%) of both locations generating similar levels of output. Such a result was expected though, given the close geographical proximity of the two locations already mentioned in Section \ref{sub:res_magnitude}. In addition, using a conversion technology unable to harness frequently occurring high wind speeds in Greenland (due to relatively low cut-out speeds, see Section \ref{subsub:average_load_factor} for detailed discussion) can further justify increased proportions of simultaneously occurring low-output events in both Greenlandic locations compared to the all-European case, which is translated into a high value of the $C_{11}$ coefficient. The two remaining cases assessing the effects of aggregating European and Greenlandic locations show contrasting results. Looking at the joint assessment of the two onshore generation sites (FR-GR$\textsubscript{on}$), one sees a fairly even distribution of occurrences across the four possible bins and a rather high share of simultaneously low-output occurrences (22\%) in both locations, an aspect that can be attributed to (i) the use of a sub-optimal conversion technology in Greenland and (ii) a relatively poor wind resource associated to the European location. In opposition, the aggregation of the two offshore locations (DK-GR$\textsubscript{off}$) reveals a very good score for high output in at least one of the locations (91\%), a result that supports the high quality wind potential suggested in Section \ref{sub:res_magnitude}, as well as the lack of correlation between wind regimes.

\begin{table}
	\centering
	\renewcommand{\arraystretch}{1.2}
	\caption{Complementarity factors $c^{(l_1,l_2)}_{nm}$ for each pair $(l_1,l_2)$ of considered locations, considering a capacity factor threshold ($\alpha$) of 30\%. Two wind speed thresholds define two different classes for low - \textbf{1} - (below $v_{min}^{\alpha}$ and above $v_{max}^{\alpha}$) and  high - \textbf{2} - (between $v_{min}^{\alpha}$ and $v_{max}^{\alpha}$) power output for a given conversion technology.}
	\label{tab:corr}
	\renewcommand\arraystretch{1.5}
	\resizebox{0.8\columnwidth}{!}{%
		\begin{tabular}{cccc}
			\multicolumn{2}{c}{} & \multicolumn{2}{c}{\textbf{FR}}\\
			\multicolumn{2}{c}{} & \textbf{1} & \textbf{2} \\ \cline{2-4}
			\multirow{2}{*}{\textbf{DK}} & \textbf{1} & 0.17 & 0.13 \\
			& \textbf{2} & 0.36 & 0.34 \\
			\cline{2-4}
		\end{tabular}
		
		\begin{tabular}{cccc}
			\multicolumn{2}{c}{} & \multicolumn{2}{c}{\textbf{GR\textsubscript{off}}}\\
			\multicolumn{2}{c}{} & \textbf{1} & \textbf{2} \\ \cline{2-4}
			\multirow{2}{*}{\textbf{GR\textsubscript{on}}} & \textbf{1} & 0.23 & 0.19 \\
			& \textbf{2} & 0.07 & 0.51 \\
			\cline{2-4}
		\end{tabular}
	}
	\\\vspace{0.5cm}
	\resizebox{0.8\columnwidth}{!}{%
		\begin{tabular}{cccc}
			\multicolumn{2}{c}{} & \multicolumn{2}{c}{\textbf{GR\textsubscript{on}}}\\
			\multicolumn{2}{c}{} & \textbf{1} & \textbf{2} \\ \cline{2-4}
			\multirow{2}{*}{\textbf{FR}} & \textbf{1} & 0.22 & 0.31 \\
			& \textbf{2} & 0.20 & 0.27 \\
			\cline{2-4}
		\end{tabular}
	
		\begin{tabular}{cccc}
			\multicolumn{2}{c}{} & \multicolumn{2}{c}{\textbf{GR\textsubscript{off}}}\\
			\multicolumn{2}{c}{} & \textbf{1} & \textbf{2} \\ \cline{2-4}
			\multirow{2}{*}{\textbf{DK}} & \textbf{1} & 0.09 & 0.21 \\
			& \textbf{2} & 0.21 & 0.49 \\
			\cline{2-4}
		\end{tabular}
	}
\end{table}

\begin{figure}
	\centering
	\includegraphics[width=0.49\textwidth]{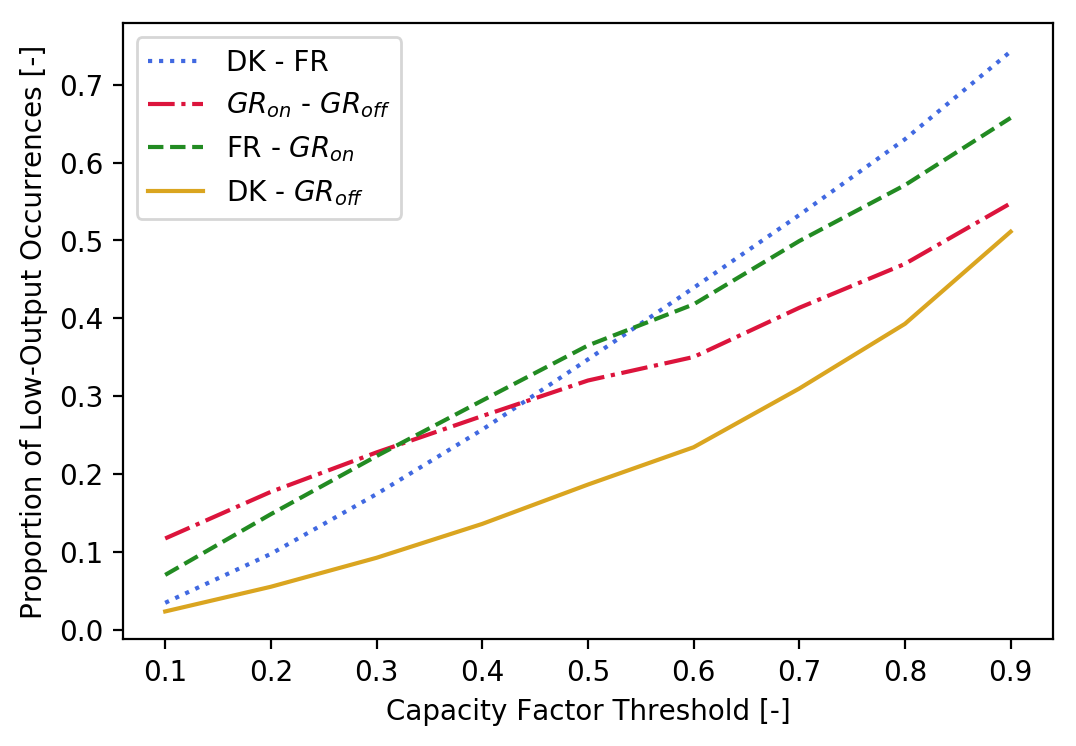}
	\caption{Proportion of low-output occurrences against various capacity factor thresholds ($\alpha$) for a given conversion technology.}
	\label{fig:complementarity}
\end{figure}

When defining the concept of complementarity, the emphasis was placed on the occurrence of detrimental low-generation events across systems. Indeed, when analysing complementarity factors as in Table \ref{tab:corr}, we are mostly interested in assessing simultaneous occurrences of low power output (that is, the $C_{11}$ element of the complementarity matrices above) for a given location pair. In this regard, Figure \ref{fig:complementarity} displays the evolution of the $C_{11}$ score for each considered location pair against different capacity factor threshold values. First, for the aggregation of the two European locations (DK-FR), a linear increase in the proportion of low-output events is observed as the capacity factor threshold increases. Next, the close geographical proximity (and, thus, the highly correlated resource) of the two Greenlandic locations (GR$\textsubscript{on}$-GR$\textsubscript{off}$) results in relatively high $C_{11}$ values for low capacity factor thresholds. For larger values of the latter, the influence of superior wind resource leads to a milder increase in low-generation events probability compared to the three other cases. Considering the FR-GR$\textsubscript{on}$ case, inferior resource associated with the European onshore node and a sub-optimal use of the conversion technology in the onshore Greenlandic location lead to higher shares of low-output events compared to the aggregation of European locations, for capacity factor thresholds smaller than 55\%. Above this threshold, the two curves intersect, driven mainly by superior wind resource in GR$\textsubscript{on}$ with respect to DK. By far, the lowest occurrences of low-generation events for any capacity factor threshold considered is associated with the aggregation of the two offshore locations (DK-GR$\textsubscript{off}$), which are characterized by high-quality and uncorrelated wind regimes.

\subsection{Critical Time Windows Analysis}
\label{sub:res_critical}

The influence of Greenlandic locations on the ($\delta$, $\alpha$)-critical time windows outcome is detailed in Table \ref{tab:windows} for six capacity factor threshold levels (i.e., from 20\% to 70\%) and four time window lengths (i.e., one hour, six hours, one day and one week). As expected, for any location pair considered, the proportion of critical time windows increases as the capacity factor threshold $\alpha$ increases. However, for each location pair, there exists a capacity factor threshold at which the evolution of the criticality index with respect to the time window length $\delta$ changes. That is, for low values of $\alpha$, the criticality index is inversely proportional with the length of the time window, while for higher values of the capacity factor threshold, the trend reverses and the two become proportional. This behaviour stems from (i) the relative position of $\alpha$ with respect to the average capacity factor observed at each location pair and (ii) the definition of the $N_{\delta}$ measure in Section \ref{subsub:critical_time_window_occurence} as a mapping returning the average of its argument. For example, the two Greenlandic locations have an aggregated average capacity factor of 55\%. On the one hand, for capacity factor thresholds below this limit, the probability of a time window to be critical decreases with $\delta$ since low-generation events (relative to $\alpha$) have a stronger impact on shorter time windows. For larger values of $\delta$, such events are often neutralized through the averaging $N_{\delta}$ mapping. On the other hand, for capacity factor thresholds above 55\%, the $\Gamma_{\delta\alpha}$ score increases proportionally with $\delta$, since less frequent high-production events (that may otherwise render shorter time windows not critical) are cancelled out over longer time windows via the measure mapping $N_{\delta}$.

\begin{table}[!b]
	\begin{center}
		\renewcommand{\arraystretch}{1.2}
		\caption{Values of $\Gamma_{\delta \alpha}$ computed from the intersection of only the two European sites (black), the intersection of only the Greenland sites (\textbf{\textcolor{ForestGreen}{green}}) and for the intersection of all four locations (\textit{\textcolor{NavyBlue}{blue}}).}
		\label{tab:windows}
		\begin{tabular}{c|c|c|c|c|c|c}
			\diagbox{$\delta$}{$\alpha$} & $20 \%$ & $30 \%$ & $40 \%$ & $50 \%$ & $60 \%$ & $70 \%$ \\
			\hline
			\multirow{ 3}{*}{1} & 0.11 & 0.18 & 0.27 & 0.35 & 0.44 & 0.53 \\
			& \textbf{\textcolor{ForestGreen}{0.14}} & \textbf{\textcolor{ForestGreen}{0.19}} & \textbf{\textcolor{ForestGreen}{0.25}} & \textbf{\textcolor{ForestGreen}{0.30}} & \textbf{\textcolor{ForestGreen}{0.35}} & \textbf{\textcolor{ForestGreen}{0.42}}          \\
			& \textit{\textcolor{NavyBlue}{0.02}} & \textit{\textcolor{NavyBlue}{0.04}} & \textit{\textcolor{NavyBlue}{0.07}} & \textit{\textcolor{NavyBlue}{0.11}} & \textit{\textcolor{NavyBlue}{0.17}} & \textit{\textcolor{NavyBlue}{0.23}}          \\
			\hline  
			\multirow{ 3}{*}{6} & 0.10 & 0.18 & 0.27 & 0.35 & 0.45 & 0.54 \\
			& \textbf{\textcolor{ForestGreen}{0.12}} & \textbf{\textcolor{ForestGreen}{0.18}} & \textbf{\textcolor{ForestGreen}{0.24}} & \textbf{\textcolor{ForestGreen}{0.30}} & \textbf{\textcolor{ForestGreen}{0.37}} & \textbf{\textcolor{ForestGreen}{0.44}}               \\
			& \textit{\textcolor{NavyBlue}{0.01}} & \textit{\textcolor{NavyBlue}{0.04}} & \textit{\textcolor{NavyBlue}{0.07}} & \textit{\textcolor{NavyBlue}{0.11}} & \textit{\textcolor{NavyBlue}{0.17}} & \textit{\textcolor{NavyBlue}{0.25}}          \\
			\hline
			\multirow{ 3}{*}{24} & 0.08 & 0.16 & 0.26 & 0.36 & 0.48 & 0.59	\\
			& \textbf{\textcolor{ForestGreen}{0.06}} & \textbf{\textcolor{ForestGreen}{0.12}} & \textbf{\textcolor{ForestGreen}{0.19}} & \textbf{\textcolor{ForestGreen}{0.28}} & \textbf{\textcolor{ForestGreen}{0.39}} & \textbf{\textcolor{ForestGreen}{0.53}}             \\
			& \textit{\textcolor{NavyBlue}{0.01}} & \textit{\textcolor{NavyBlue}{0.02}} & \textit{\textcolor{NavyBlue}{0.06}} & \textit{\textcolor{NavyBlue}{0.11}}& \textit{\textcolor{NavyBlue}{0.20}} & \textit{\textcolor{NavyBlue}{0.32}}          \\
			\hline
			\multirow{ 3}{*}{168} & 0.01 & 0.06 & 0.18 & 0.38  & 0.58 & 0.77  \\ 
			& \textbf{\textcolor{ForestGreen}{0.00}} & \textbf{\textcolor{ForestGreen}{0.01}} & \textbf{\textcolor{ForestGreen}{0.06}} & \textbf{\textcolor{ForestGreen}{0.18}} & \textbf{\textcolor{ForestGreen}{0.43}} & \textbf{\textcolor{ForestGreen}{0.75}}                \\
			& \textit{\textcolor{NavyBlue}{0.00}} & \textit{\textcolor{NavyBlue}{0.00}} & \textit{\textcolor{NavyBlue}{0.01}} & \textit{\textcolor{NavyBlue}{0.08}} & \textit{\textcolor{NavyBlue}{0.26}} & \textit{\textcolor{NavyBlue}{0.58}} \\
		\end{tabular}
	\end{center}
\end{table}

An interesting result in Table \ref{tab:windows} concerns the lower proportion of critical time windows corresponding solely to the sites in Greenland (in green) and for generation thresholds above 30\% compared to the same values related to the European locations (in black). Selected sites in Greenland are in close geographical proximity and this feature attracts strong non-complementarity in terms of air mass dynamics. However, lower values compared to the aggregation of European locations suggest better wind potential due to the existence of more constant local wind flows, i.e., the katabatic winds. Moreover, for short time window lengths (one day at most) the gain in terms of critical time window occurrence (i.e., the difference between $\Gamma_{\delta\alpha}$ values associated with the same ($\delta$, $\alpha$) parameters) increases as the generation threshold grows, again indicating superior wind magnitudes associated with the sites in Greenland. Finally, the advantage of coupling the two regions (in blue) is observed under all considered set-ups. For example, instances when wind production levels for 24-hour time windows in both Greenland and Europe drop below 70\% account for less than one third (i.e., 32\%) of the full time frame, a reduction of 27\% and 11\% compared to the Europe-only and Greenland-only cases, respectively. These results support the findings of previous sections with respect to the benefits of linking remote regions in terms of both complementarity and magnitude of wind resource.

\section{Conclusion and Future Work}
\label{sec:conclusion}
The current work evaluates Greenlandic wind resource quality through standard statistical metrics applied to raw wind data and to typical power generation proxies (i.e., capacity factors), as well as its complementarity with western European wind regimes via a systematic framework quantifying the occurrence of system-wide low-generation events. By leveraging a state-of-the-art regional climate model that has been repeatedly validated over polar regions, a promising area in southern Greenland is identified and found to exhibit vast wind power generation potential and possess complementary regimes with respect to European locations known for their high quality wind resource. These results lend further support to the claim that tapping into extensive renewable energy generation potential located in remote areas can prove beneficial for a secure and reliable supply of electricity in future power systems dominated by VRE. Another takeaway of this study pertains to the need for technological innovation in wind turbine design, a key aspect that could enable the achievement of even higher capacity factors in Greenlandic regions swept by high quality wind resource.

Regarding further research directions, analysis of wind regimes at different heights above ground level is of considerable interest taking into account the particular features of Greenlandic katabatic flows. In this regard, increased average capacity factors are anticipated at lower elevations (e.g., 50 metres above ground level), where the increased influence of topography and heat transfer processes bolsters a more frequent occurrence of semi-permanent katabatic flows, while the cut-out speeds of wind converters are reached less often. Another assessment path consists in developing a tailored analysis to quantify the potential benefits of a Greenlandic wind farm supplying Europe through an HVDC interconnection. In addition, a mapping of various regulatory (e.g., investment mechanisms, remuneration schemes, operational and trading features) and geopolitical aspects is envisioned in order to provide a more complete view of the complexity surrounding the development of interconnectors as part of a global electricity network.

\bibliography{katabatic}

\end{document}